# Gamma Standing Wave in the Photonic Crystal of Resonant Rh Nuclei

**Y Cheng and B Xia**

Department of Engineering Physics, Tsinghua University, Beijing 100084, P. R. China
Email: yao@tsinghua.edu.cn

**Abstract.** In a previous report, we have shown that the rhodium lattice consisting of resonant nuclei is an ideal photonic crystal in nature. Plenty of extraordinary observations are attributed to the collective down conversion of the multipolar nuclear transition; in particular the spontaneous open up of photonic band gap. Emissions of directionality depending on the macroscopic geometry manifest that the standing wave is global in the polycrystalline sample. In this work, further observations of the directional emissions are summarized. By applying an external magnetic field at room temperature, not only the predicted macroscopic nuclear polarization but also its strong directionality are demonstrated. The standing wave lasts for more than hundred hours at room temperature in the single crystal despite its natural half-life of one hour. The so-called nuclear Raman Effect between two M4 transitions of $^{193m}$Ir and $^{195m}$Pt and the E3 transition of $^{103m}$Rh is discovered, which gives the brand new aspects to detect gravitational waves.

## 1. Introduction

The nuclear exciton of $^{103m}$Rh in photonic crystal (PC) consisting of resonant Rh nuclei is a critical phenomenon depending on (i) inversion density, (ii) temperature, (iii) earth gravity, (iv) sample geometry and (v) magnetic field. The former four are reported previously [1]. In this work, the predicted macroscopic nuclear polarization [1] in polycrystal is demonstrated by an external dc magnetic field. In analogy with the atomic transition near the band edge within photonic band-gap (PBG) material [2], the photon localized near nuclei is a photon-nucleus bound state with a steady and fractional nuclear inversion. The collective decay channels of cascaded down-conversion [1] are mainly distributed near the band edge characterized by the long localization length [2]. The localized superradiance is thus spreading in the whole crystal as a global standing wave with a vanishing group velocity even in polycrystal. Investigation on the single crystal shows that the standing wave survives for more than one hundred hours, though the characteristic emissions of $^{103m}$Rh have decreased below the background noise within twenty hours after the bremsstrahlung irradiation. The standing wave with particular spatial symmetry at lattice sites provides the so-called nuclear Raman Effect, when the impurity is radioactive of appropriate transitions. The impurity becomes an active sensor to detect the standing wave inside the crystal. Moreover, this long-lasting standing wave allows the sample cooling down to a very low temperature preferred by the highly sensitive measurement of gravitational waves (GWs). The applied rhodium samples are polycrystal and single crystal of 99.9% and 99.999% $^{103}$Rh, respectively. One of $10^{12}$ $^{103}$Rh nuclei is excited to be $^{103m}$Rh by the 120-min bremsstrahlung irradiation from a 6-MeV linear accelerator [3].

John and his collaborators [2] have predicted the macroscopic polarization of the atomic steady state for a highly inverted initial population with small initial polarization. Though initial inversion and initial polarization are extremely low in our studies, we have shown some macroscopic features of the $^{103m}$Rh exciton state to suggest the nuclear polarization in [1]. Here, we demonstrate the macroscopic nuclear polarization by applying an external magnetic field. The applied magnetic field is less than 15 Gauss, which shall not give any significant nuclear magnetic susceptibility at room temperature in the conventional sense. However, the polarization of the photon-nucleus bound state is *rotated* and probably *mode-switched* between orders of nuclear polarization by the applied dc magnetic field. Spontaneous cascade down-conversion (SCDC) is the branching channel to catalyze the long-lived Mössbauer decay [1]. The SCDC spectrum peaked near the half transition energy (20 keV) manifests that the biphoton cascade with the similar single-photon decay has the major contribution [4]. Cascade photon pairs ought to be m-beam nuclear Borrmann modes (mNBMs) [5] of the opposite parity. The energy-time entangled [6] SCDC pair is bounded near nuclei [2] in the particular direction of the most significant nuclear Borrmann channels (NBCs) [5] dictated by lattice, gravity, temperature distribution and macroscopic sample geometry. The 3D fcc lattice of rhodium becomes 1D-like PC with the matched symmetry between lattice and multipolar transition. Superradiance and exciton diffusion as a function of temperature reveal the long-lived Mössbauer Effect of atto-eV linewidth [1]. In the conventional sense, this extremely narrow linewidth shall be destroyed by the temperature variation inside the sample [7]. New physics of the long-lived Mössbauer Effect emerges in the PC of identical resonant nuclei. SCDC entanglement is analogous to the entangled Cooper pair in superconductor. Nuclear resonant absorption of the SCDC γs is preserved by PBG to inhibit the thermal agitation in analog with the superconducting gap [1]. Exciton states pinched at the *air defect* [2] of impurities are analogous to the color centers of semiconductor and the pinning center in superconductor [1]. The local intensity of standing wave at impurity sites is depleted by the inelastic scattering, *i.e.* color centers dissipate the standing wave being a sensor to detect the standing wave inside the crystal. When the impurity is a radioactive element, color center is no longer passive but active.

The SCDC entanglement creates two particular types of inelastic scatterings with ionization energies greater than 40 keV, *i.e.* rhodium K hypersatellites and the high-Z impurity K lines [1]. Nuclei of $^{103m}$Rh belong to the category of active color center, whereas the high-Z impurities at ground state are passive color centers. The inelastic scatterings of Pt and Bi K lines are shown in this work. Of particular interest is the Bi binding energy greater than 80 keV, *i.e.* twice the transition energy of $^{103m}$Rh. Two types of color centers have different evolution in time, *i.e.* active color centers depend on the inversion population and the intensity of standing wave, while the passive color centers only depend on the intensity of standing wave.

Due to the dependences on (i) temperature, (ii) collecting angle and (iii) sample geometry, counts of energy between 41 keV and 360 keV are denoted by bunching photons in [1]. By inserting filter sheets, we identify that the so-called bunching photons in the energy range above 100 keV are consisting of relative high-energy photons such as the high-Z K lines rather than the Rh K lines. During the series studies of 6 days, two active color centers of 10.53-day $^{193m}$Ir and the 4.02-day $^{195m}$Pt are accumulated day by day. Counts of bunching photons do not depend on the accumulation during the sequence of studies. Bunching photons are thus attributed to the atomic transitions of impurities instead of their nuclear emissions such as the gammas of $^{193m}$Ir and $^{195m}$Pt. Strange directional bunching photons is attributed to the SCDC superradiance, which has been directly observed by cooling [1]. Inelastic scatterings of random distributed impurities coherently driven by the standing wave shall not give the collective emissions of directionality, unless they are entangled multi-photons in NBC due to the similar mechanism to generate the directional SCDC propagation.

The standing wave is excited to the maximal intensity immediately after the 2-hour bremsstrahlung irradiation, *i.e.* bremsstrahlung is in favor of the standing wave. Relaxation of the standing wave gives the initial decay of impurity K lines. Standing wave becomes steady and lasts to the subsequent irradiation despite the $^{103m}$Rh decay. Counts of impurity K lines and bunching photons do not depend on the irradiation sequence, *i.e.* the standing wave is not accumulated and reset during the bremsstrahlung irradiation. The stationary emissions from the passive color centers indicate the depleted local photon intensity near color centers and balanced by the diffusion rate of nuclear exciton. The diffusion rate increases by means of the liquid-nitrogen cooling to give immediate response of passive color centers, while the response of active rhodium nuclei are delayed due to the non-Markovian radiation reservoir [1,2]. Due to fewer defects, the standing wave is enhanced in the single crystal. When the Rh K counts have decayed below the background noise at twentieth hour after irradiation, three revivals [8] of Rh K lines subsequently occurs. Third but final revival shows that the standing wave is steady at least for 130 hours. The standing wave is strongly coupled with Rh nuclei in the extreme high-Q cavity of PC, *i.e.* the decay of standing wave is slower than the isolated $^{103m}$Rh [8]. Nuclear resonant absorption manifests the long-lived Mössbauer effect.

Nuclear Raman Effect has been suggested to catalyzing the multipolar nuclear transitions via virtual states induced by photon or graviton with the conserved symmetry [1,9]. To have significant response from weak GWs, the Raman scattering of "graviton in and photon out" requires collective amplification [9]. Instead of observing the fast superradiance directly, the color centers provide the other possibility to observe the stationary standing wave in crystal. The nuclear Raman Effect has been observed by the intensive bremsstrahlung irradiation [10]. Strongly coupled defect modes in PC provide the speed-up decay of $^{193m}$Ir and $^{195m}$Pt at the low radiation intensity. Of particular concern are the collective nuclear transitions of distributed $^{193m}$Ir, $^{195m}$Pt and $^{103m}$Rh together. This observation reported in this work reveals the potential applications of γ lasing and GW detection. Doping of the appropriate long-lived nuclei may become a sensitive GW detector in the low-frequency band of periods extended to weeks.

## 2. Experimental setup

The major experimental setup and the irradiation procedure are the same as reported in the case-I study of [1]. Two orientations of A and C of the central irradiation spot in [1] are investigated (figure 1). For the C orientation, one rhodium sample of 1mm×25mm×25mm is applied. For the A orientation, three rhodium samples sandwiched by 0.5-mm tungsten and 0.05-mm plastic sheets are stacked together such that the 25-mm axis is toward the detector. The magnetic field is 7.62±0.01 Gauss/A given by the Helmholtz coil in the N-S horizontal direction toward the detector. The earth magnetic field in the horizontal direction is minimized at 0.04 A. The power supply has the peak-to-peak ripple current of 10 mA at the line harmonics of 50 Hz without significant dependence on output dc current (Table 1). The investigation on the (100) oriented single crystal (Goodfellow, RH002110, ϕ 8mm, 1-mm thickness, 99.999% purity) applies a different detector system (Coaxial GE detector of CANBERRA EGC-45-195R, CANBERRA 2002C preamplifier, 2026 amplifier and 8071 multichannel analyzer) of the upright detector with an active area of 3200mm$^2$. A low-noise lead shielding is applied for the particular measurements of single crystal. The detector system of low resolving power (~1 keV) has a large detecting angle about 1.7π sr. No external magnetic field is able to be applied except the filter of 35-μm copper sheets in [1]. During the measurement, the sample is placed atop the upright detector with normal vector of sample in the upright direction.

**Table 1.** Applied horizontal magnetic field in the N-S direction toward the detector (figure 1) using the Helmholtz coil of several applied dc currents (7.62±0.01 Gauss/A). Major ac components are the line harmonics with peak to peak ~10 mA. Two line harmonics are listed with arbitrary scale.

| Current (A)  | 0    | 0.04 | 0.125 | 0.25 | 0.5 | 1   | 2    |
|--------------|------|------|-------|------|-----|-----|------|
| Field (Gauss)| -0.3 | 0.0  | 0.65  | 1.6  | 3.5 | 7.3 | 14.9 |
| 50 Hz        | 5.2  | 4.8  | 5.5   | 5.5  | 5.0 | 5.2 | 4.4  |
| 100 Hz       | 1.0  | 1.3  | 0.8   | 0.7  | 0.8 | 0.6 | 0.7  |

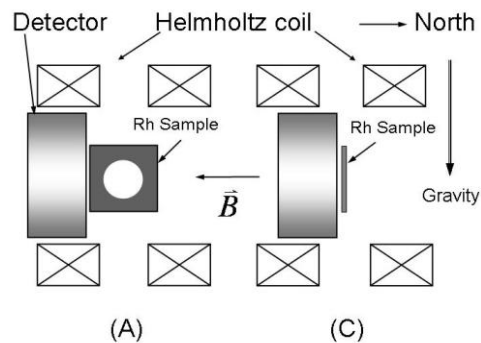

**Figure 1.** Set up in two orientations A and C. The irradiation spot locates at the sample center for all studies reported in this work.

## 3. Spectral Profiles

Figures 2-9 illustrate the time evolution of spectral deformations $\tilde{S}_i(\omega,t)$ deviated from the normal profiles $\bar{S}_i(\omega)$. The measured spectra $S_i(\omega,t)$ are normalized by individual total counts, channel by channel in time

$$\tilde{S}_i(\omega,t) = A_i \left( \frac{S_i(\omega,t)}{\int S_i(\omega,t) d\omega} - \bar{S}_i(\omega) \right) \quad (1)$$

with factors $A_i$ chosen for the presentation of each band for figures. The subscript $i$ stands for Kα and Kβ and γ. Factors $A_i$ are selected in accordance with reference [1].

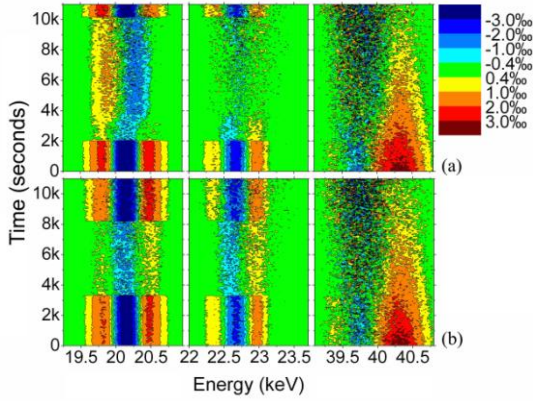

**Figure 2.** Spectral deformations of Kα, Kβ, and γ under an external magnetic field in orientation C of (a) 0.04 A; (b) 0.125 A. Total counts in each band are normalized with $A_{K\alpha}=1$, $A_{K\beta}=0.5$ and $A_\gamma=0.25$ as defined in eq. (1). The measuring time is three hours. Decay at 40.3 keV is the peak pile-up of rhodium Kα. Revival of the H1 phase [1] is observed in these two measurements.

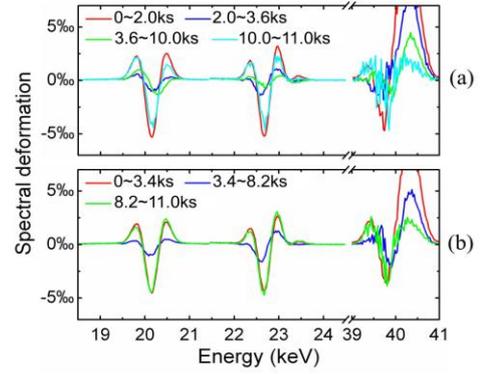

**Figure 3.** Averaged spectral deformations of Kα, Kβ, and γ corresponding to each phase illustrated in figure 1 except all $A_i=1$. Of particular concern is the Kβ hypersatellites at 23.2 keV, which are significant during the H1 phase [1]. Due to the low Kα pile-up at the H1 revival, the PBG of H1 phase at 40 keV becomes free from the pile-up count but restricted by the resolving power (~ 0.4 keV) of the detecting system.

### 3.1. Orientation C

We show four studies in orientation C under the supplied current of 0.04 A, 0.125 A, 0.5 A and 2 A. Revivals of the H1 phase defined in [1] within three hours are observed in two studies of 0.04 A and 0.125 A. Revival of the H1 phase is absent for ten hours in two studies of 0.5 A and 2 A. The Kβ$_{1,3}$ hypersatellites located at 23.2 keV are significant in the H1 phase (figures 3 and 5), which reveals the multiple ionization of two K holes. Deformations at the H1 phase are enhanced by the ac component of the external magnetic field, whereas the SCDC counts at 21.5 keV are less enhanced (table 2 and detailed later in the section 4). The deformation amplitudes of four studies reveal the main superradiance along the long sample edge is *rotated* to the weak superradiance in the normal direction of sample. Of particular concern is the simultaneous drop of K and γ counts in figure 4b, which will be detailed later on in the section 6.

**Table 2.** ΔKα and SCDC counts normalized by Kα count of the H1 phase in figures 3 and 5.

| Current (A) | 0 | 0.04 | 0.125 | 0.5 | 2 |
|---|---|---|---|---|---|
| $\left|\tilde{S}_{K\alpha}(20.2keV,H1)\right|$ (10$^{-3}$) | 2.16±0.07 | 5.30±0.06 | 4.51±0.05 | 6.57±0.05 | 4.20±0.05 |
| $\tilde{S}_{K\alpha}(21.5keV,H1)$ (10$^{-4}$) | 1.18±0.02 | 1.16±0.02 | 1.14±0.02 | 1.17±0.02 | 1.16±0.02 |

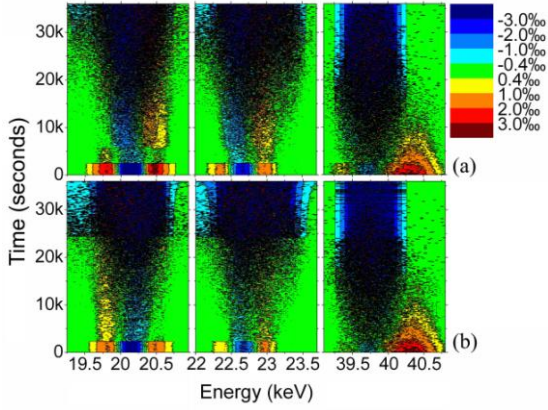
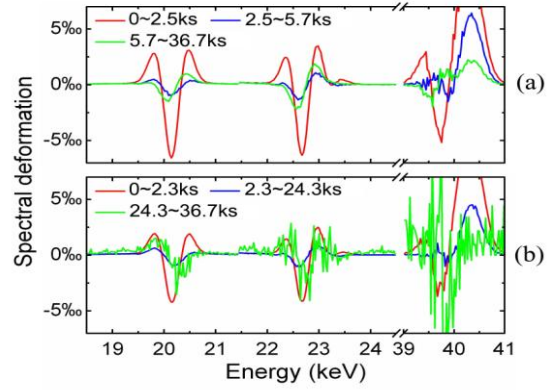

**Figure 4.** Spectral deformations of Kα, Kβ, and Δγ under an external magnetic field in orientation C of (a) 0.5 A; (b) 2 A. Total counts in each band are normalized with $A_{K\alpha}=1$, $A_{K\beta}=0.5$ and $A_{\gamma}=0.25$ as defined in eq. (1). The measuring time is ten hours.

**Figure 5.** Averaged spectral deformations of Kα, Kβ, and γ corresponding to each phase illustrated in figure 3 except all $A_i=1$. No revival of the H1 phase is observed in ten hours. The counts of Kα, Kβ, and γ at 24.3 ks in (b) are suddenly reduced together with [195m]Pt counts shown in figures 15-16.

*3.2. Orientation A*

One typical spectral deformation of six studies in orientation A is shown in figures 6 and 7. Applied currents are 0.04A, 0.125A, 0.25A, 0.5A, 1A and 2A (table 3a). All measurements are performed for three hours except the study of 2A for ten hours. There are in general no clear phase transitions for the studies in orientation A due to the fact that three separated samples provide the mixing emissions from three exciton states. The H1 phase appears at the first 4400 s of the 2-A study shown in figures 6 and 7, though the irradiation spot is far from the sample edge.

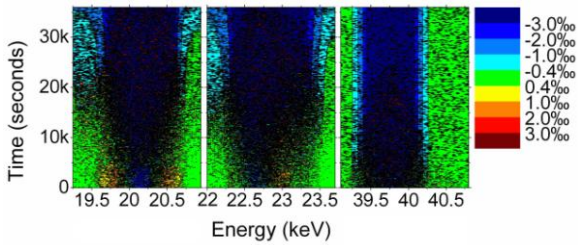
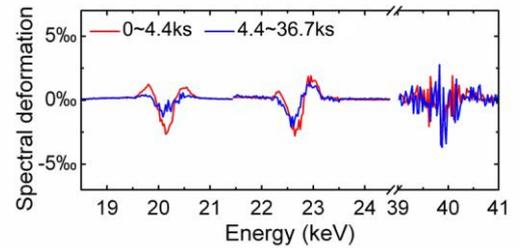

**Figure 6.** Spectral deformations of Kα, Kβ, and γ under external magnetic field in orientation A of the 2-A measurement. Total counts in each band are normalized with $A_{K\alpha}=1$, $A_{K\beta}=0.5$ and $A_{\gamma}=0.25$ as defined in eq. (1).

**Figure 7.** Averaged spectral deformations of Kα, Kβ, and γ corresponding to each phase illustrated in figure 7 except all $A_i=1$. No revival of the H1 phase is observed in ten hours.

*3.3. Single Crystal*

Figures 8 and 9 show the time evolution of spectral deformations of single crystal with the particular setup detailed in section 2. Two typical results of measurements are shown. One of them is done by inserting a 35-μm copper sheet between the sample and the detector. The slight baseline shift and the pile up of the applied analog system have been removed by calibration. No phase transition is observed.

Of particular concern are the different spectral deformations between two cases. The Kα lines overlap the Kβ lines due to the low detector resolution about 1 keV that leads to the stepwise discontinuity at 21.5 keV between two individually normalized bands of Kα and Kβ. Jumps in figure 9b are constant, while jumps in figure 9a are time-dependent. Discontinuity is mainly contributed by the time-dependent spectral deformation and partially contributed by the decaying SCDC buried in Kα and Kβ peaks [1]. The K deformation of inserting copper filter in figure 9(b) is less than the deformation without filter in figure 9(a). Same behavior has been observed in [1], but treated as different phases. The cascade decay of two K holes creates the energy-time entanglement between hypersatellite and its successive decay. The K deformations and the SCDC counts are entangled biphoton of great similarity in the joint biphoton absorption [11], which occurs on the copper sheet by simultaneous absorption of two noninteracting atoms at different locations. The enhancement of biphoton absorption is a non-classical effect ascribed to the biphoton nature of the energy-time entanglement [6]. When the signal and the idler photons are absorbed by the inserting copper sheet together, the joint probability of biphoton absorption as a function of the degree of biphoton entanglement [11] is greater than the probability to absorb two independent photons.

The FWHM resolution at 40 keV is 1.02±0.003 keV, while the FWHM resolution of 40-keV joint biphoton detection shall be higher than 1.02 keV ascribed to the bias of the multi-channel analyzer. We observe the γ distribution getting narrower down to 0.98 keV in figure 8. Deformation of γ is thus changing from the PBG dip to the central pick due to the fact that the measured profile is getting narrower than the calibrated profile. Similar observation will be shown in the next section, where the central dip evolves to the central pick resolved by much higher spectral resolution.

We extend the spectral profiles to demonstrate the insignificant sidebands at 17.4 keV, 18.3 keV and 24.3 keV.

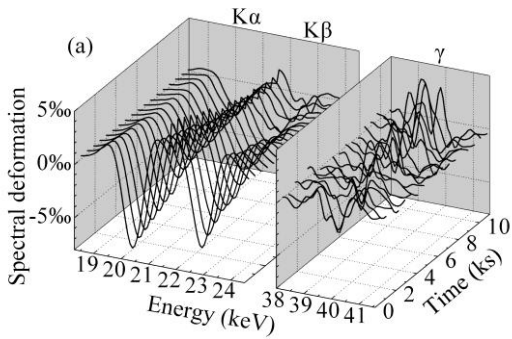

**Figure 8.** Time evolution of the spectral deformations of Kα and Kβ from single crystal with total count $A_i$ =1 for three bands without filter. Each line is the ten-minute accumulation of data in three hours.

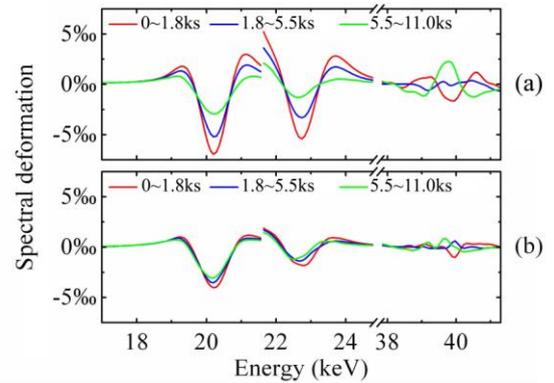

**Figure 9.** Spectral deformations of Kα and Kβ from single crystal with total count $A_i$ =1 for three bands of (a) no filter; (b) 35-μm copper filter. Since there is no phase transition, the separation of time section is arbitrarily selected.

*3.4. Photonic Band Gap*

The PBG profile at 39.76 keV is affected by the Kα2 pile-up at 40.3 keV. Only on the condition of low Kα count rate in figures 3 and 5, the PBG profiles are free from the pile-up count, in particular during the H1 revival in figure 3. We show the PBG profile of various phases in figures 9 and 10 by removing the pile-up contribution using the normal profile $\bar{S}_{K\alpha}(\omega)$ of (1). The error of corrected PBG values in figure 6 is less than 10 ppm. Figures 10 and 11 demonstrate the combined profiles between PBG and detector characteristic corresponding to each phase.

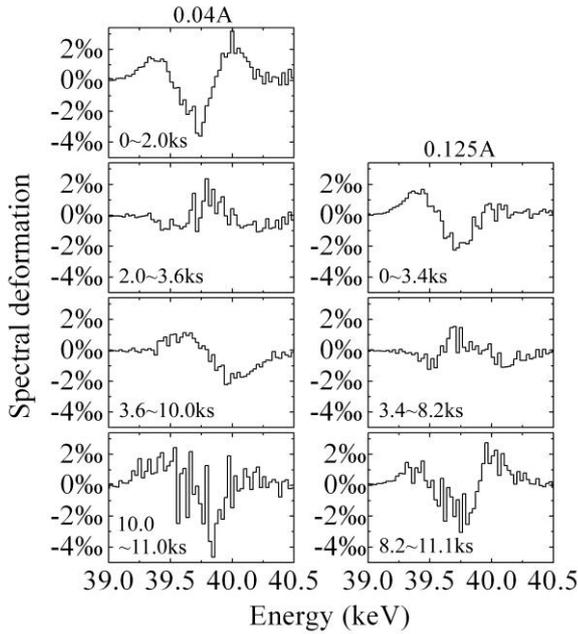

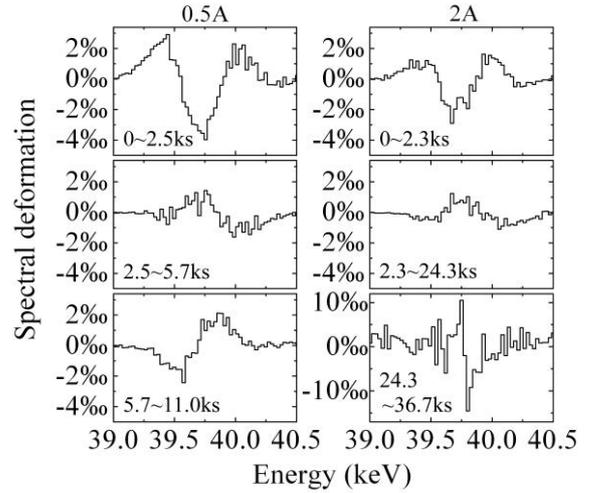

**Figure 10.** γ deformation as the combined profiles between PBG and detector characteristic corresponding to each phase in figure 2 with the external magnetic field of 0.04 A and 0.125 A.

**Figure 11.** γ deformation as the combined profiles between PBG and detector characteristic corresponding to each phase in figure 4 with the external magnetic field of 0.5 A and 2 A.

## 4. Macroscopic nuclear polarization

Macroscopic nuclear polarization of Rh nuclei in polycrystal is predicted by the global localized superradiance [1,2]. We apply a magnetic field of 15 Gauss or less. No susceptibility of Rh nuclei shall be observed at room temperature, unless the predicted macroscopic polarization is available. Though the irradiation dosages are same, K and γ counts are manipulated by the magnetic field shown in figure 12 and tables 3a and 3b. Significant dips of photon counts are observed at 0.04 A when the horizontal magnetic field toward the detector is absent. The other dips appear at 0.5 A probably due to the *mode-switching* between orders of nuclear polarization.

The K conversion $\alpha_K$ is the measured K count divided by the measured γ count. In fact that the Borrmann transmitted superradiance is mainly along the long sample edge [1], $\alpha_K$ in orientation C is higher than $\alpha_K$ in orientation A. Counts of two SCDC bands, cγ1 (14-19 keV) and cγ2 (24-38 keV), roughly follow the K and γ counts, respectively. Their ratios in table 3 demonstrate the effect of magnetic field, particularly in orientation A. Large escape of γ indicates the biphoton existence [1]. Counts of bunching photons with significant effect of magnetic field in orientation A indicate the incorporation of polarization with superradiance. In orientation C, the polarization is slightly *rotated away* from the major direction such that the magnetic effect is weak (also in table 2). Sidebands located at 6.4 keV and distributed near 24.3 keV are found in addition to the SCDC sideband at 17.4 keV [4]. Three sidebands illustrated in figure 13 are drastically enhanced in orientation A, particularly the sideband at 17.4 keV. These sidebands are insignificant in single crystal (see figure 9) thus probably due to impurities or grain boundaries scattered by the SCDC superradiance.

**Table 3a.** K, γ, K conversion $\alpha_K$, bunching photons (BP) of energy between 100 keV and 210 keV, cγ1/Kα SCDC (14-19 keV) normalized by Kα count, cγ2/γ SCDC (24-38 keV) normalized by γ count, γe/γ escape of γ at 30 keV normalized by the γ count and normalized counts of three sidebands (6.4 keV/Kα, 17.4 keV/Kα, 24.3 keV/γ) depending on the applying magnetic field. Three-hour data are measured in orientation A.

| Current (A) | 0 | 0.04 | 0.125 | 0.25 | 0.5 | 1 | 2 |
|---|---|---|---|---|---|---|---|
| K ($10^4$) | 185.3±0.1 | 219.4±0.1 | 232.1±0.2 | 243.8±0.2 | 216.0±0.2 | 246.9±0.2 | 276.6±0.2 |
| γ ($10^4$) | 1.64±0.01 | 1.93±0.01 | 2.07±0.02 | 2.17±0.02 | 1.91±0.01 | 2.19±0.02 | 2.46±0.02 |
| $\alpha_K$ | 112.2±0.9 | 112.9±0.8 | 111.5±0.8 | 111.3±0.8 | 112.5±0.8 | 111.8±0.8 | 111.4±0.7 |
| BP($10^3$) | 2.1±0.2 | 1.1±0.2 | 6.0±0.2 | 5.8±0.2 | 3.0±0.2 | 5.4±0.2 | 4.9±0.2 |
| cγ1/Kα ($10^{-2}$) | 2.75±0.01 | 2.73±0.01 | 2.68±0.01 | 2.69±0.01 | 2.72±0.01 | 2.83±0.01 | 2.79±0.01 |
| cγ2/γ ($10^{-1}$) | 3.01±0.07 | 2.82±0.06 | 3.23±0.006 | 3.04±0.06 | 2.99±0.06 | 3.31±0.06 | 3.12±0.05 |
| γe/γ ($10^{-2}$) | 2.1±0.2 | 2.0±0.2 | 2.0±0.2 | 2.1±0.2 | 2.1±0.2 | 2.1±0.2 | 2.0±0.2 |
| 6.4 keV/Kα ($10^{-4}$) | 8.8±0.4 | 8.4±0.4 | 5.4±0.4 | 6.2±0.4 | 7.6±0.4 | 7.5±0.4 | 7.1±0.4 |
| 17.4 keV/Kα ($10^{-4}$) | 15.0±0.7 | 15.7±0.4 | 10.5±0.4 | 10.4±0.4 | 15.8±0.4 | 11.9±0.4 | 14.8±0.4 |
| 24.3 keV/γ ($10^{-2}$) | 0.9±0.2 | 1.0±0.2 | 1.4±0.2 | 0.6±0.2 | 1.0±0.2 | 1.1±0.2 | 1.0±0.1 |

**Table 3b.** Three-hour data are measured in orientation C.

| Current (A) | 0 | 0.04 | 0.125 | 0.5 | 2 |
|---|---|---|---|---|---|
| K ($10^4$) | 5751±0.8 | 5524±0.7 | 5717±0.8 | 5538±0.7 | 5830±0.8 |
| γ ($10^4$) | 50.01±0.08 | 47.97±0.07 | 49.74±0.08 | 48.02±0.07 | 50.84±0.08 |
| $\alpha_K$ | 115.0±0.2 | 115.1±0.2 | 114.9±0.2 | 115.3±0.2 | 114.8±0.2 |
| BP ($10^3$) | 0.80±0.04 | 2.2±0.05 | 2.2±0.05 | 2.1±0.05 | 2.1±0.05 |
| cγ1/Kα ($10^{-2}$) | 2.536±0.003 | 2.580±0.003 | 2.594±0.003 | 2.586±0.003 | 2.594±0.002 |
| cγ2/γ ($10^{-1}$) | 2.95±0.01 | 2.93±0.01 | 2.92±0.01 | 2.94±0.01 | 2.92±0.01 |
| γe/γ ($10^{-2}$) | 2.12±0.03 | 2.20±0.03 | 2.16±0.03 | 2.18±0.03 | 2.14±0.03 |
| 6.4 keV/Kα ($10^{-4}$) | 1.90±0.06 | 1.83±0.06 | 1.75±0.06 | 1.88±0.06 | 1.81±0.06 |
| 17.4 keV/Kα ($10^{-4}$) | 0.10±0.07 | 0.23±0.06 | 0.11±0.06 | 0.05±0.06 | 0.12±0.06 |
| 24.3 keV/γ ($10^{-2}$) | 0.15±0.03 | 0.21±0.02 | 0.22±0.02 | 0.14±0.02 | 0.23±0.02 |

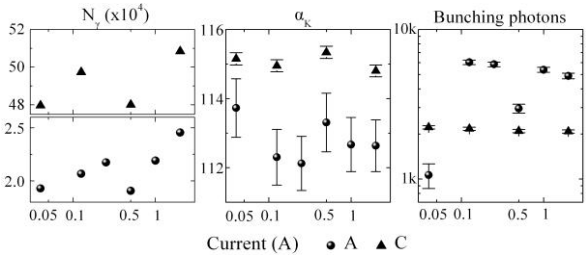

**Figure 12.** γ counts $N_\gamma$, K conversion $\alpha_K$, and bunching photons of energy between 100 keV and 210 keV in three hours depending on the external magnetic field. The $^{195m}$Pt γ at 129.7 keV has been removed from the count of bunching photons. Dots are the data in the sample orientation A and triangles are the data in orientation C.

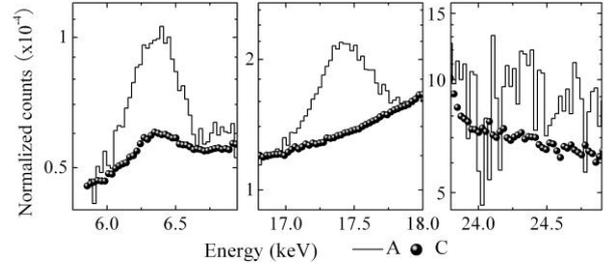

**Figure 13.** Three sidebands at 6.4 keV, 17.4 keV and 24.3 keV accumulated by six measurements in orientation A and four measurements in orientation C listed in table 3. Three reported sidebands may consist of several peaks. Dots are the data in the sample orientation A and triangles are the data in orientation C.

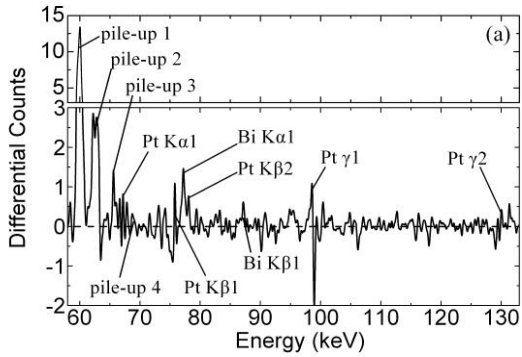

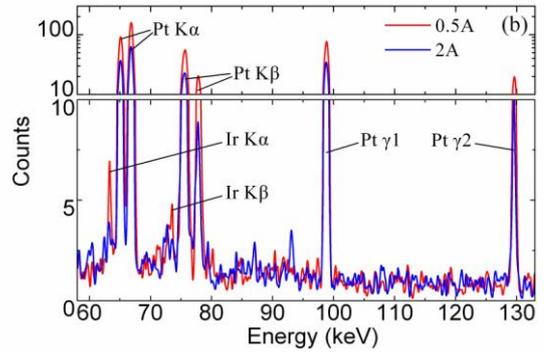

**Figure 14.** Spectral profiles of the impurity color centers measured in orientation C. (a): Differential counts between the first 20 minutes and the successive 20 minutes of figure 4b (2A). Four pile-ups are the double or the triple pile-up of six combinations among Rh γ at 39.76 keV, Rh Kα at 20.2 keV and Rh Kβ at 22.7 keV respectively. Pt γ1 (98.9 keV) and Pt γ2 (129.7 keV) are emissions from $^{195m}$Pt. (b): Spectral profiles are accumulated in the late 204 minutes of the ten-hour measurements of 0.5 A and 2 A, respectively.

## 5. Color centers of impurities

Figures 14a demonstrates the impurity K lines attributed to the inelastic scattering of standing wave, of which amplitude reaches the maximum after the 120-minute irradiation. Platinum K lines and bismuth K lines are given by the differential counts due to the relaxation of standing wave. The $^{195m}$Pt of 4.02-day half-life is excited by the bremsstrahlung irradiation reported previously [1], while the $^{193m}$Ir of 10.53-day half-life is reported for the first time. We shall not observe any differential contribution of $^{195m}$Pt and $^{193m}$Ir between the first 20 minutes and the successive 20 minutes due to their long lifetimes of days. Thus, the differential residuals of Pt K lines in figure 14a are contributed by the inelastic scattering.

Figure 14b shows two spectral profiles, which are accumulated during the late 204 minutes from two ten-hour studies of 0.5 A and 2 A, respectively. While the $^{195m}$Pt can be identified by Pt γ1 and Pt γ2, the $^{193m}$Ir transition is not able to be identified by the associated γ at 80.24 keV due to its weak intensity. The absent Ir K lines in figure 14a reveal that the observed Ir K lines in figure 14b are caused by the $^{193m}$Ir conversion. Since the 2-A measurement is performed after the 0.5-A measurement, $^{193m}$Ir and $^{195m}$Pt of the 2-A measurement are accumulated more than that of 0.5 A. After the sudden reduction in counts at the 397$^{th}$ minute (see figures 4b, 15, and 16), figure 14b shows the reduced emissions from $^{193m}$Ir and $^{195m}$Pt together. Ratios between Pt K lines and Pt γs in figure 14b is 20% higher than the theoretical value of conversion rates, the partial K lines are contributed by the inelastic scattering as shown in figure 14a.

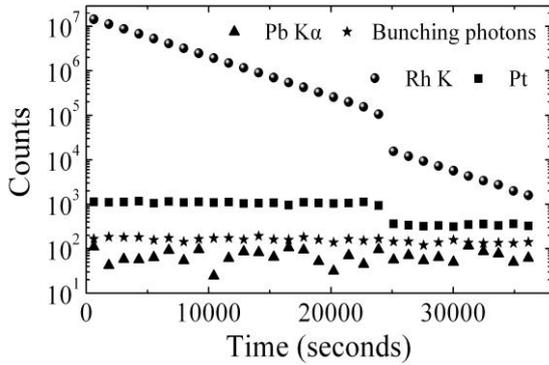

**Figure 15.** Time evolution of various emissions from $^{103m}$Rh, $^{195m}$Pt, bunching photons and lead shielding of the figure 3b. The shown data are collected per twenty minutes. The $^{103m}$Rh counts consist of Kα and Kβ from rhodium. The Pt counts consist of Kα, Kβ, γ1 at 98.9 keV and γ2 at 129.7 keV. Bunching photons are collected from 100 keV to 210 keV of removed background and the $^{195m}$Pt γ2 at 129.7 keV.

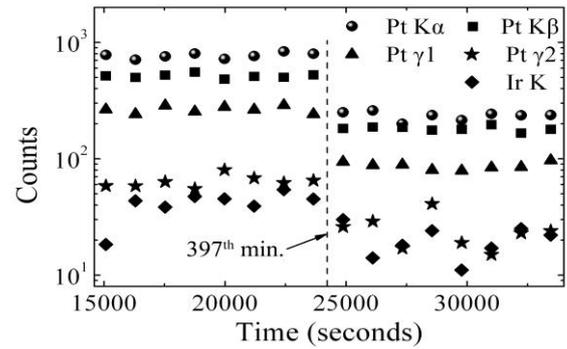

**Figure 16.** Time evolution of four $^{195m}$Pt emissions of Kα, Kβ, γ1 at 98.9 keV, γ2 129.7 keV and Ir K lines from $^{193m}$Ir before and after the suddenly reduced counts at the specific time of the 397$^{th}$ minute in figure 14. The shown data are accumulated for twenty minutes.

## 6. Nuclear Raman Effect

Figure 15 illustrates the time evolution of various emissions from $^{103m}$Rh, $^{195m}$Pt, bunching photons and lead shielding of the figure 4b. The stationary counts from lead shielding demonstrate the detecting system is in the normal operating condition and the sudden reduction of the emissions from $^{103m}$Rh and $^{195m}$Pt at the 397$^{th}$ minute of measurement is not due to any artifact. Figure 16 shows the time evolution of the emissions from $^{193m}$Ir and $^{195m}$Pt before and after the specific time of the simultaneously count reduction.

Lattice site of the rhodium crystal is the nodal point of the standing waves featured by the $^{103m}$Rh E3 transition with particular spatial distribution. When rhodium atom is replaced by $^{193m}$Ir or $^{195m}$Pt, the impurity nuclei *see* the same field distribution as emitted from the Rh nuclei due to the invariant translation under lattice constants. In the air defects of $^{193m}$Ir or $^{195m}$Pt, active nuclei of the M4 transition are strongly coupled to the virtual states of opposite parity with spin change ΔJ≤3 induced by the E3 standing wave. One part of the virtual states provides the E1 transition to catalyze the M4 transition and *vice versa*.

The reported 2-A measurement happens to be the last measurement in the 6-day period of experiment. Active color centers of $^{193m}$Ir and $^{195m}$Pt have been accumulated to the highest inversion population after 16 sequential bremsstrahlung irradiations. Figure 17 shows that the collective transitions at the 397$^{th}$ minute are not a spontaneous action but starts at a quarter hour before suddenly reduced counts at 23 ks, when cγ2 count starts to increase. We summarize following observations after the 397$^{th}$ minute of measurement (i) bunching photons remain at the same level; (ii) counts of γ and cγ2 increase and oscillate at relative high level; (iii) K count slightly oscillates; (iv) lost of

$^{103m}$Rh is significantly greater than $^{193m}$Ir and $^{195m}$Pt; (v) Kα/Kβ ratio drops from 3.41±0.01 to 3.21±0.04, and (vi) the number of stepwise $^{103m}$Rh lost is about $10^{10}$.

Of particular concern is the simultaneous interaction among three different transitions. No coupling between defect modes is expected for the active impurities of different nuclear transition energies with distributed locations in sample. The reported collective speed-up decays require further investigation.

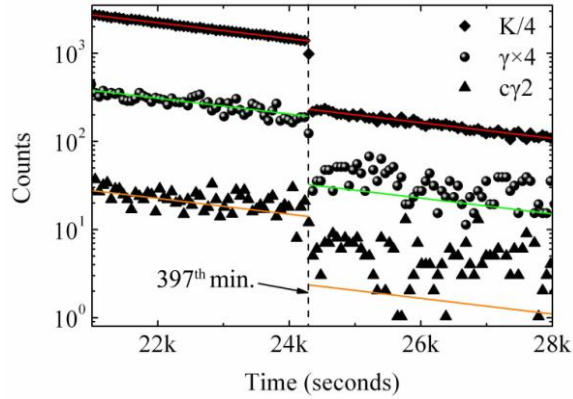

**Figure 17.** (Color online) Time evolution of K, γ and cγ2 in every minute before and after the suddenly reduced count at the specific time of the 397$^{th}$ minute in figure 14. Lines are the natural decay scaling by the ratios of $α_K$=114.8 and cγ2/γ=0.0292 listed in table 2 from K count as the inversion density. Ratios of $α_K$ and cγ2/γ in the shown time period are 118±2, 0.33±0.01 before drop and 77±4, 0.60±0.04 after count reduction, while the count ratio of Kα/Kβ is reduced from the normal value of 3.41±0.01 to 3.21±0.04.

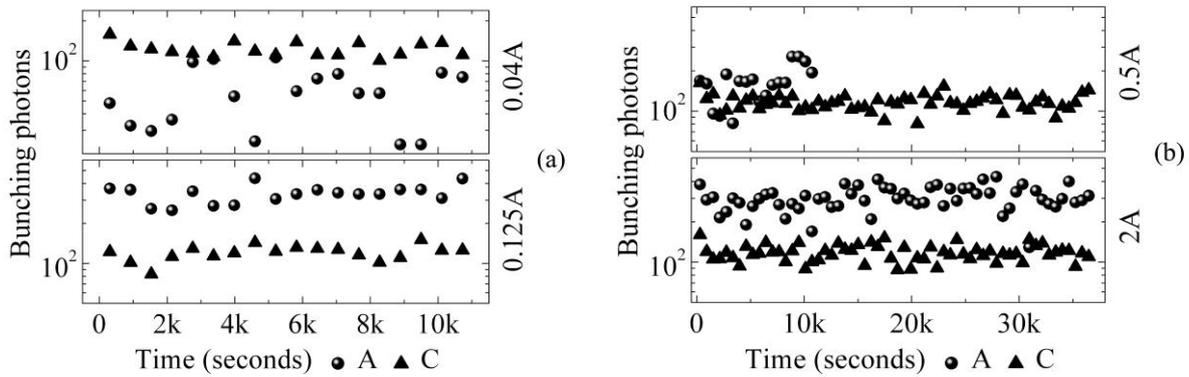

**Figure 18.** Time evolution of bunching photons of collective energies between 100 keV and 210 keV in the sample orientations A and C, (a) three-hour measurements of 0.04 A and 0.125 A; (b) ten-hour measurements of 0.5 A and 2 A, except the 0.5-A measurement in orientation A is done in three hours. The contribution of $^{195m}$Pt γ2 at 129.7 keV has been removed. Each data point is the count accumulated for ten minutes. Dots are the data in the sample orientation A and triangles are the data in orientation C.

## 7. Steady Bunching Photons

Though no direct evidence by means of coincide measurement is available, counts above 100 keV are assumed to be bunching photons due to several reasons, *i.e.* (i) no atomic transition is higher than 100 keV; (ii) counts depend on temperature [1], sample geometry, and magnetic field (figure 12); (iii) counts depend on the collecting angle; (iv) spectral profile does not match the profile of detector efficiency; (v) counts do not depend on the inversion populations of $^{103m}$Rh, $^{193m}$Ir and $^{195m}$Pt; and (vi) counts do not depend on the irradiation sequence, *i.e.* independent to the accumulation of other long-lived states. All of the studies show the decay behavior of bunching photons occurring in the first twenty minutes. Therefore, they are related to the SCDC superradiance without significant contribution from the nuclear Raman Effect of $^{193m}$Ir and $^{195m}$Pt. Figures 18a and 18b demonstrate the typical time evolution of bunching photons, which is steady and last for ten hours. Result of longer measuring time will be reported in the next section. The counts of bunching photons are reduced, when the power supply is turned off at 0 A (Table 3). Thus, the ac component of supplied current promotes the bunching photons. Reduction of bunching photons at 0.04A and strong dependence on magnetic field in orientation A indicates that the major SCDC biphoton superradiance mainly consists of transitions of ΔJ=3.

All counts of bunching photons in orientation A are higher than that in orientation C, when the sample surfaces of emission are taken into account. We have shown the entangled biphoton absorption of the cascade Rh K lines of multiple ionizations by the external absorption sheet in figure 9. On the contrast, the penetration of the cascade Rh K lines inside the crystal is tenfold enhanced by cooling in orientation A [1]. The directionality of bunching photons suggests that they are due to multiple ionizations and attributed to the same mechanism observed by the cascade Rh K lines, *i.e.* entangled propagation in NBC.

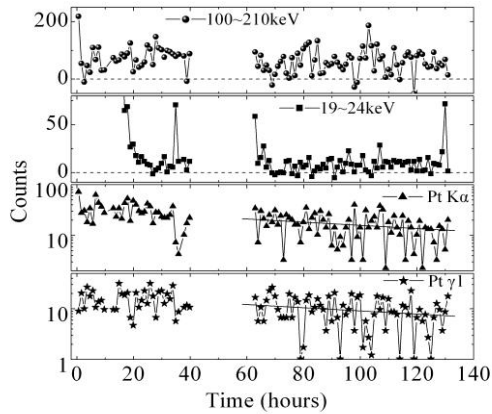

**Figure 19.** Time evolution of bunching photons, Rh K lines, Pt Kα and $^{195m}$Pt γ1 at 98.9 keV from the single crystal. The lines at two Pt emissions indicate the decay speed of 4.02-day half-life. Standard deviations of 100-210 keV, 19-24 keV, Pt Kα, and Pt γ1 are 39, 6, 11, and 6, respectively.

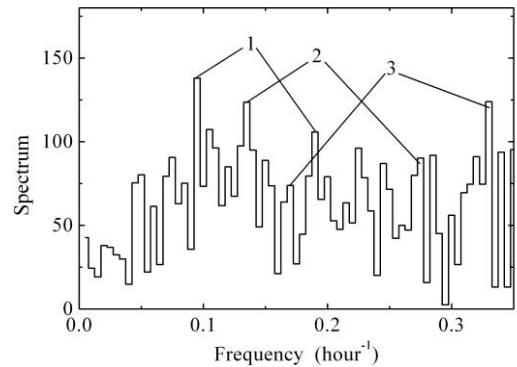

**Figure 20.** Spectrum of the $^{195m}$Pt γ1 at 98.9 keV shown in figure 19. Three harmonic pairs of particular concern are located at (1): 0.095-0.19 hr$^{-1}$; (2): 0.135-0.27 hr$^{-1}$; and (3): 0.165-0.33 hr$^{-1}$. This measurement starts at 9:51 PM, 2006, 6, 13, Beijing time.

## 8. Gamma standing wave in single crystal

Though the single crystal is highly purified, we observe significant emissions from $^{195m}$Pt. Of particular interest are the following observations in figure 19. Firstly, the nuclear Raman Effect is demonstrated by the γ1 at 98.9 keV of $^{195m}$Pt, which does not follow the decay of 4.02-day half life for the first forty hours. Secondly, three sequential revivals of Rh K lines are observed. The last one occurs at the 130$^{th}$ hour after irradiation. Later revivals are expected. All of the numbers of K revival obtained at 35$^{th}$, 63$^{rd}$ and 130$^{th}$ hour is ten-times greater than the standard deviation of the background noise. Thirdly, bunching photons are steady for 130 hours, which are especially selected between 100 keV and 210 keV in accordance with the numbers of the polycrystal sample listed in table 3. Similar stationary time evolution is extended to the energy region higher than 500 keV.

We observe the overly large fluctuations in figure 19. It is interesting to know whether the fluctuation of γ1 at 98.9 keV has some characteristic frequencies. Though the temperature of sample changes by day and night, the 24-hour modulation is absent in figure 20. The tidal frequency of 12.4 hour is also absent, but three peaks of harmonic pairs located at frequencies of 0.095-0.19 hr$^{-1}$, 0.135-0.27 hr$^{-1}$ and 0.165-0.33 hr$^{-1}$.

## 9. Conclusion

Revival of the Rh K lines reveals the strong coupling between nuclei and standing wave in the structured reservoir consisting of the individual high-Q nuclear cavities on lattice sites. The standing wave of the localized superradiance survives in the single crystal for more than hundred hours, *i.e.* longer than the natural lifetime of isolated $^{103m}$Rh by two orders of magnitude. When the active color centers such as $^{195m}$Pt are doped into the rhodium crystal, the bremsstrahlung can excite $^{103m}$Rh and $^{195m}$Pt together. Interaction between two long-lasting states greatly extends the frequency range of the potential GW detection.

Color centers of impurities locate at the nodal point of the standing wave with particular spatial distribution emitted from $^{103m}$Rh. When the color center is a radioactive element of appropriate transitions in the inverted rhodium crystal, the nuclear Raman Effect occurs at the *air defect* of PC despite the low radiation intensity. The nuclear Raman Effect to speed up the decay of three transitions of $^{103m}$Rh, $^{193m}$Ir and $^{195m}$Pt together is observed. Similar speed-up decay of $^{103m}$Rh has been observed with $^{195m}$Pt together (rediscovered in [12]) at the moment of the liquid-nitrogen quenching, which is considered to be the stimulated emission. Moreover, the observed nuclear Raman Effect in single crystal lasts for forty hours under the earth magnetic field at room temperature.

We have shown the global photonic state of standing wave depending on the macroscopic sample geometry [1]. The directional emissions with geometric dependence are evident under an external magnetic field. Field-dependent observations reveal the macroscopic nuclear polarization at room temperature. Two significant sidebands of SCDC superradiance at 6.4 keV and 24.3 keV are identified in addition to the sideband at 17.4 keV reported previously [1]. They have several features (i) weakly depending on the external magnetic field; (ii) strongly depending on the sample geometry; and (iii) insignificant in the single crystal. These sidebands require further investigation to give us more information on the details of the long-lasting standing wave inside the crystal.

K lines and their bunching emissions from color centers either passive or active are significantly enhanced by the ac component of applied magnetic field. Counts of bunching photons do not depend on the inversion population of

$^{103m}$Rh and its exciton phase, but decay in the beginning of measurements and are enhanced by cooling. Depletion of the standing wave locally at the high-Z color centers is manifested.

## 10. Statement of the main author

Plenty of discoveries are first realized after the kindly experimental support has been stopped. We give our best to detail the relevant observations by the documentary-note style of this work.

## 11. Acknowledgements

We thank Yi-Kang Pu for the proofreading of manuscript. YC thanks the support of Tipei Li and Yexi He. This work is supported by the NSFC grant 10675068.

## 12. References

[1] Cheng Y, Xia B 2007 arXiv: 0706.0960v2

[2] John S and Quang T 1995 *Phys. Rev. Lett.* **74** 3419; John S and Quang T 1994 *Phys. Rev. A* **50** 1764; Vats N and John S 1998 *Phys. Rev. B* **58** 4168; Woldeyohannes M and John S 2003 *J. Opt. B: Quantum Semiclass. Opt.* **5** R43

[3] Cheng Y, Xia B, Liu Y-N, Jin Q-X 2005 *Chin. Phys. Lett.* **23** 2530; Cheng Y, *et al.* 2006 *Hyperfine Interactions* **167** 833

[4] Enaki N A and Macovei M 1997 *Phys. Rev. A* **56** 3274; Enaki N A and Mihalache D 1997 *Hyperfine Interactions* **107** 333; Enaki N A 1988 *JETP* **67** 2033

[5] Hutton J T, Hannon J P and Trammell G T 1988 *Phys. Rev. A* **37** 4269; Hannon J P and Trammell G T 1999 *Hyperfine Interactions* **123**/**4** 127

[6] Shih Y H 2003 *IEEE Journal of selected Topics in Quantum Electronics* **9** 1455

[7] Josephson B D 1960 *Phys. Rev. Lett.* **4** 341; Pound R V and Rebka G A 1960 *Phys. Rev. Lett.* **4** 274

[8] Lambropoulos P, Nikolopoulos G M, Nielsen T R and Bay S 2000 *Rep. Prog. Phys.* **63** 455

[9] Cheng Y and Shen J Q 2007 arXiv:gc-qc/0703066

[10] Collins C B, Eberhard C D, Glesener J W and Anderson J A 1988 *Phys. Rev. C* **37**, 2267; Carroll J J, Byrd M J, Taylor K N, Richmond D G, Sinor T W, Hodge W L, Paiss Y, Eberhard C D, Anderson J A, Collins C B, Scarbrough E C, Antich P P, Agee F J, Davis D, Huttlin G A, Kerris K G, Litzand M S, Wittaker D A 1991 *Phys. Rev. C* **43** 1238

[11] Muthukrishnan A, Agarwal G S and Scully M O 2004 *Phys. Rev. Lett.* **93** 093002

[12] Cheng Y and Wang ZM 2006 arXiv:quant-ph/0611031